\documentclass[aps,twocolumn,showpacs,float]{revtex4}

\usepackage{amsmath}
\usepackage{graphicx}
\usepackage{float}
\floatstyle{boxed}
\usepackage{bm}

\begin{document}

\bibliographystyle{prsty}
\author{Eugene M. Chudnovsky$^{1}$ and Reem Jaafar$^2$}
\affiliation{$^{1}$Department of Physics and Astronomy, Lehman College and Graduate School, The City University of New York, 250 Bedford Park Boulevard West, Bronx, NY 10468-1589\\
$^{2}$Department of Mathematics, Engineering and Computer Science, LaGuardia Community College, The City University of New York, 31-10 Thomson Avenue, Long Island City, NY 11101}
\date{\today}

\begin{abstract}
We show that the magnetic moment of a nanoparticle embedded in the surface of a solid can be switched by surface acoustic waves (SAW) in the GHz frequency range via a universal mechanism that does not depend on the structure of the particle and the structure of the substrate. It is based upon generation of the effective ac magnetic field in the coordinate frame of the nanoparticle by  the shear deformation of the surface due to SAW. The magnetization reversal occurs via a consecutive absorption of surface phonons of the controlled variable frequency. We derive analytical equations governing this process and solve them numerically for the practical range of parameters. 
\end{abstract}

\title{Manipulating Magnetization of a Nanomagnet by Surface Acoustic Waves: Spin-Rotation Mechanism}

\maketitle

Switching of the magnetic moment by means other than applying the magnetic field has been one of the paradigms of modern magnetism. Studies of the spin transfer torque have led to the commercialization of random access memory (STT-RAM) devices \cite{Brataas-NatureMaterials2012}. Recent years have been also marked by intensive research on manipulating magnetic moments by electric fields in multiferroic and composite materials \cite{Matsukura-NatureNano2015}. As far as the speed is concerned, a direct $180$-degree switching of the magnetization by the electric field would be the most desirable for applications.  This approach, however, requires significant ingenuity because  linear coupling of the magnetic moment to the electric field is prohibited by symmetry. 

\begin{figure}
\vspace{-1cm}
\includegraphics[width=85mm]{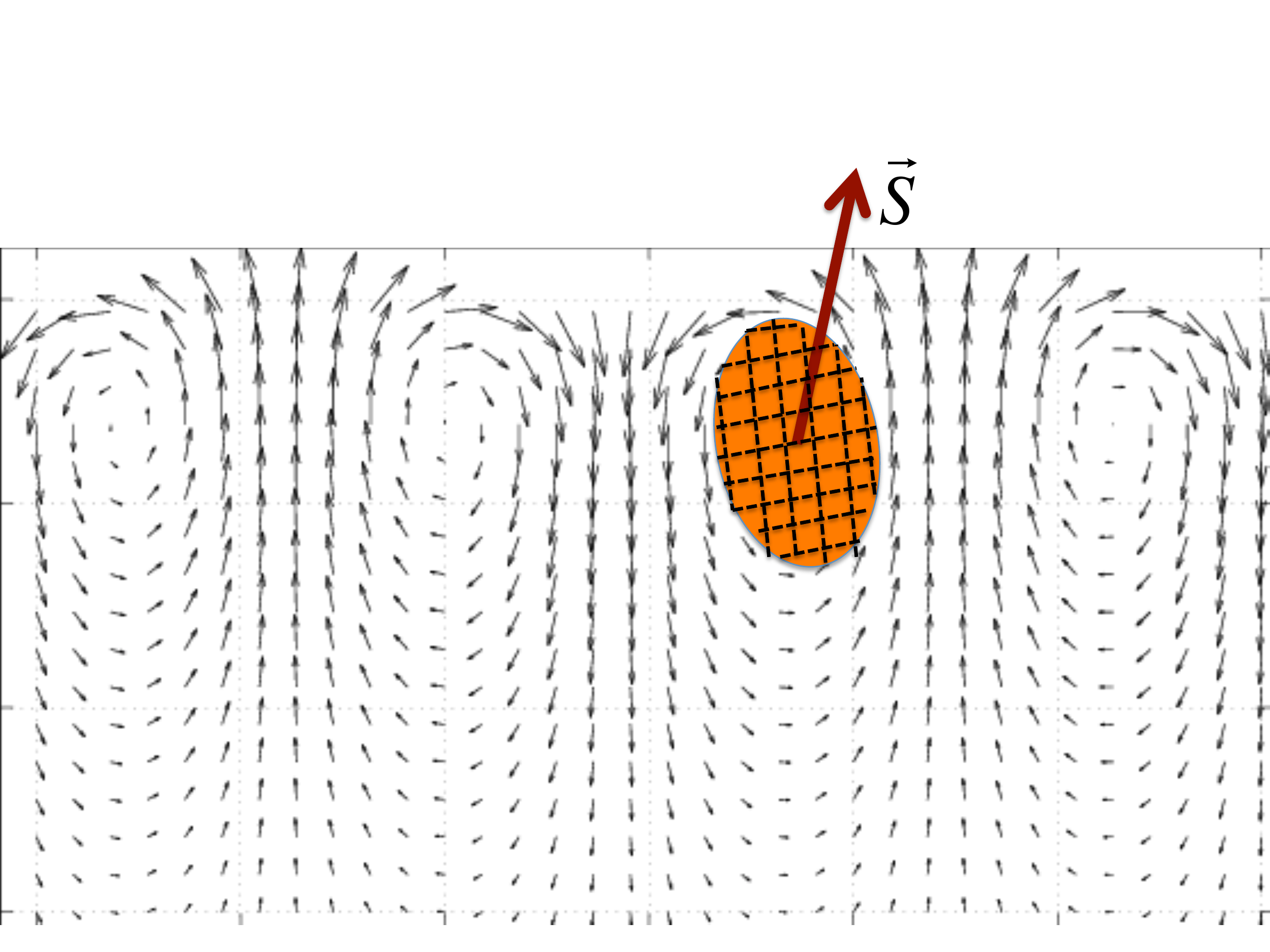}
\caption{Color online: Schematic representation of the spin-rotation coupling provided by SAW. A magnetic nanoparticle with a total spin ${\bf S}$ is adhered to a solid surface. Elastic displacements due to SAW (shown by small black arrows of decreasing length as one goes down away from the surface) generate fast rotational motion of the nanoparticle. In the coordinate frame of the particle the rotations produce the effective ac magnetic field that reverses its magnetic moment via pumping of energy and angular momentum.}
\label{SAWwithSpin}
\end{figure}

In this letter, we propose switching of the magnetic moment of a nanoparticle, embedded in a solid surface, by mechanical oscillations due to surface acoustic waves (SAW). Magnetization dynamics due to magnetostriction induced by SAW in a ferromagnetic layer has been studied before \cite{Davis-APL2010,Kovalenko-PRL2013,Thevenard-PRB2013,Thevenard-PRB2014,Davis-JAP2015}. The spin-rotation mechanism proposed here is based upon observation that a nanoparticle subjected to SAW undergoes rotational oscillations with the angular velocity ${\bm \Omega} = \frac{1}{2} {\bm \nabla} \times \dot{\bf u}$, where ${\bf u}$ is the local displacement field, see Fig. \ref{SAWwithSpin}. In the rotating coordinate frame of the nanoparticle its magnetic moment ${\bf M}$ feels the effective ac magnetic field, ${\bf h}_{ac} = {\bm \Omega}/\gamma$, with $\gamma$ being the gyromagnetic ratio. This provides the linear coupling between ${\bf M}$ and ${\bm \Omega}$. We believe that this effect has been been partially responsible for the spin dynamics generated by SAW in manganites \cite{Macia-PRB07} and molecular magnets \cite{Macia-PRB08}. 

The spin-rotation coupling responsible for the above mechanism is a manifestation of the Einstein - de Haas effect at the nanoscale \cite{chugarsch-PRB2005,garchu-PRB2015}. It corresponds to the transfer of the angular momentum between spin and mechanical degrees freedom. In that sense it relies on the spin-orbit interaction that manifests itself in the crystal field acting on the magnetic moment. Interaction of SAW with the magnetic moment is provided by the rotation of the crystal field. One interesting feature of this mechanism is that it depends only on the frequency of the ferromagnetic resonance and on the switching field which can be independently measured. Spin-rotation switching magnetism was previously studied in application to microcantilevers \cite{Kovalev-APL2003,Kovalev-PRL2005,Wallis-APL2006,jaachugar09prb,OKeeffe-PRB2013} and molecules sandwiched between conducting leads \cite{JCG-EPL10}. It was recently proposed for enhancing magnetization reversal by the spin polarized currents \cite{Cai-PRB2014} and for achieving magnetization reversal by a pulse of the electric field in a torsional cantilever made of a multiferroic material \cite{CJ-JAP2015}. It is well known (see, e.g., Ref. \onlinecite{CGC-PRB2013} and references therein) that the ac field of a few Oe amplitude and frequency comparable to the frequency of the ferromagnetic resonance (FMR) can reverse the magnetic moment via pumping of energy and angular momentum into the system. In quantum-mechanical terms it corresponds to a consecutive absorption of photons, each changing the $Z$-projection of the total spin ${\bf S}$ by one. In a similar manner, due to the equivalence of rotations to the magnetic field, the consecutive absorption of surface phonons can reverse the magnetic moment of a nanoparticle. 

SAW in the GHz range are easily generated by modern techniques \cite{Macia-PRB08,Li-JAP13}. Using ${\bf h}_{ac} = {\bm \Omega}/\gamma$ and ${\bm \Omega} = \frac{1}{2} {\bm \nabla} \times \dot{\bf u}$ it is easy to see that the amplitude of the effective ac field due to SAW would be of order $h_{ac} \sim \omega_{FMR}^2 u_0/(\gamma c_t)$, where $u_0$ is the amplitude of SAW and $c_t$ is the speed of the transverse sound. Consequently, at $\omega_{FMR} = 2\pi f_{FMR} \sim 10^{10}$s$^{-1}$ and $c_t \sim 10^3$m/s one only needs $u_0 \sim 10^{-10}$m = $1${\AA}  at SAW wavelength in the micrometer range to achieve $h_{ac} \sim 1\,$Oe, which is clearly practicable. We are interested in the reversing of the magnetic moments of nanoparticles of size that is small compared to the wavelength of SAW. In this case the acoustic mismatch between the particle and the medium is irrelevant, the particle is just a small inclusion that rotates together with the medium. As we shall see, by using two sources of SAW one can reproduce the key features of the magnetization dynamics achieved with a circularly polarized electromagnetic field. However, the wavelength of SAW is five orders of magnitude smaller than the wavelength of electromagnetic radiation of the same frequency, thus allowing manipulation of magnetic entities at the microscopic level. 

The transverse displacements from two sources of SAW (or more precisely, Rayleigh waves) with a $90^o$ phase shift, sending waves along $X$ and $Y$ directions, are given by \cite{LL}
\begin{eqnarray}
u_x^{(1)} & = & \kappa_t a \cos(kx -\omega t) e^{\kappa_t z},\, u_z^{(1)} = ka\sin(kx -\omega t) e^{\kappa_t z} \nonumber \\
u_y^{(2)} & = & -\kappa_t a \sin(ky -\omega t) e^{\kappa_t z},\, u_z^{(2)} = ka\cos(ky -\omega t) e^{\kappa_t z} \nonumber \\
\end{eqnarray}
where $a$ is a constant related to the amplitude of the wave, $\kappa_t =\sqrt{k^2 - {\omega^2}/{c_t^2}}$, $\omega = c_t k \xi$, and $\xi$ is the only root of the equation
\begin{equation}
\xi^6 - 8\xi^4 + 8\xi^2\left(3 - 2\frac{c_t^2}{c_l^2}\right) - 16\left(1 - \frac{c_t^2}{c_l^2}\right) = 0
\end{equation}
satisfying the condition that it is real, positive, and smaller than one. Local rotation of the solid matrix, ${\bm \varphi} = \frac{1}{2}{\bm \nabla} \times {\bf u}$, from the two sources is given by 
\begin{eqnarray}
\varphi_x & = & \frac{\partial u_z}{\partial y} - \frac{\partial u_y}{\partial z} =(\kappa_t^2 - k^2)a \sin(ky - \omega t) e^{\kappa z} \nonumber \\
\varphi_y & = & \frac{\partial u_x}{\partial z} - \frac{\partial u_z}{\partial x} = (\kappa_t^2 - k^2)a \cos(kx - \omega t) e^{\kappa z}\nonumber \\
\varphi_z & = & \frac{\partial u_y}{\partial x} - \frac{\partial u_x}{\partial y} = 0
\end{eqnarray}

We consider the nanomagnet to be small compared to the SAW wavelength. At the origin of the coordinate frame, where the magnet is located, one has
\begin{equation}
\varphi_x  =  \frac{\omega^2}{c_t^2}a \sin(\omega t), \quad  \varphi_y  =  -\frac{\omega^2}{c_t^2}a\cos(\omega t), \quad  \varphi_z  = 0
\end{equation}
In our approach $\omega$ may depend on time. The components of the angular velocity are ${\bm \Omega} = \dot{\bm \varphi}$, so that the effective ac field in the coordinate frame of the magnet is ${\bm \Omega}/\gamma$. In this connection one should notice that while the amplitude and the frequency of the electromagnetic field are independent variables, it is not so for the effective field due to rotations. We shall consider the case of constant power, $\omega u_0 = {\rm const}$, where $u_0$ is the amplitude of SAW at the surface. Since $a$ in the above equations can be written as $a = u_0/k = \xi c_t u_0/ \omega$, this corresponds to the constant amplitude of the oscillations of ${\bm \varphi}$. With that condition one obtains
\begin{eqnarray}\label{phi}
\varphi_x   & = &  \epsilon \sin(\omega t),\, \varphi_y  =  -\epsilon\cos(\omega t),\, \varphi_z  = 0  \\
\Omega_x  & = &  \epsilon(\omega + \dot{\omega}t)\cos(\omega t),\,  \Omega_y  =  \epsilon(\omega + \dot{\omega}t)\sin(\omega t),\, \Omega_z  = 0 \nonumber
\end{eqnarray}
where $\epsilon = ({\xi}/{c_t})(\omega u_0) = {\rm const} \ll 1$ is a small parameter in the problem.

In the coordinate frame of the nanomagnet placed in the magnetic field ${\bf H}$ its energy is
\begin{equation}
E_M = -KVm_z^2 - V{\bf m} \cdot {\bf H} - V{\bf m}\cdot ({\bf H} \times {\bm \varphi}) - V{\bf m} \cdot \frac{\bm \Omega}{\gamma} 
\end{equation}
where ${\bf m}$ is the magnetization and $V$ is the volume of the magnet. Here the first term is due to uniaxial magnetic anisotropy, the second and third term describe Zeeman interaction with the magnetic field whose direction in the body frame is changed by rotation, and the last term is Zeeman interaction with the effective field due to rotation in accordance with the Larmor theorem. The effective field is given by
\begin{equation}
{\bf H}_{eff} = -\frac{1}{V}\frac{\partial E_M}{\partial {\bf m}} ={\bf H}_{A} + {\bf H} + {\bf H}_\varphi 
\end{equation}
where ${\bf H}_{A} = 2Km_z\hat{\bf z}$ is the anisotropy field and 
${\bf H}_\varphi = {\bf H} \times {\bm \varphi} + \dot{\bm \varphi}/{\gamma}$
is the effective field due to rotations. When ${\bf H}$ is along the Z-axis 
\begin{equation}
{\bf H}_\varphi = \gamma^{-1}\epsilon(\gamma H + \omega + \dot{\omega}t)[\hat{\bf x}\cos(\omega t) + \hat{\bf y}\sin(\omega t)]
\end{equation}
Due to its structure it is convenient to switch (in the body frame) to another coordinate frame rotating at the angular velocity $\omega(t)$ around the Z-axis. In that frame 
\begin{equation}
{\bf H}_\varphi = \gamma^{-1}{\epsilon}(\gamma H + \omega + \dot{\omega}t)\hat{\bf x}
\end{equation}
looks in the same direction at any moment of time and the field acquires an addition $\gamma^{-1}\omega(t)\hat{\bf z}$.  The total effective field in the new coordinate frame is given by
\begin{equation}
\gamma{\bf H}_{eff} = \epsilon(\gamma H + \omega + \dot{\omega}t)a\hat{\bf x} + \left(2\gamma Km_z + \gamma H + \omega\right)\hat{\bf z}
\end{equation}

\begin{figure}
\includegraphics[width=85mm]{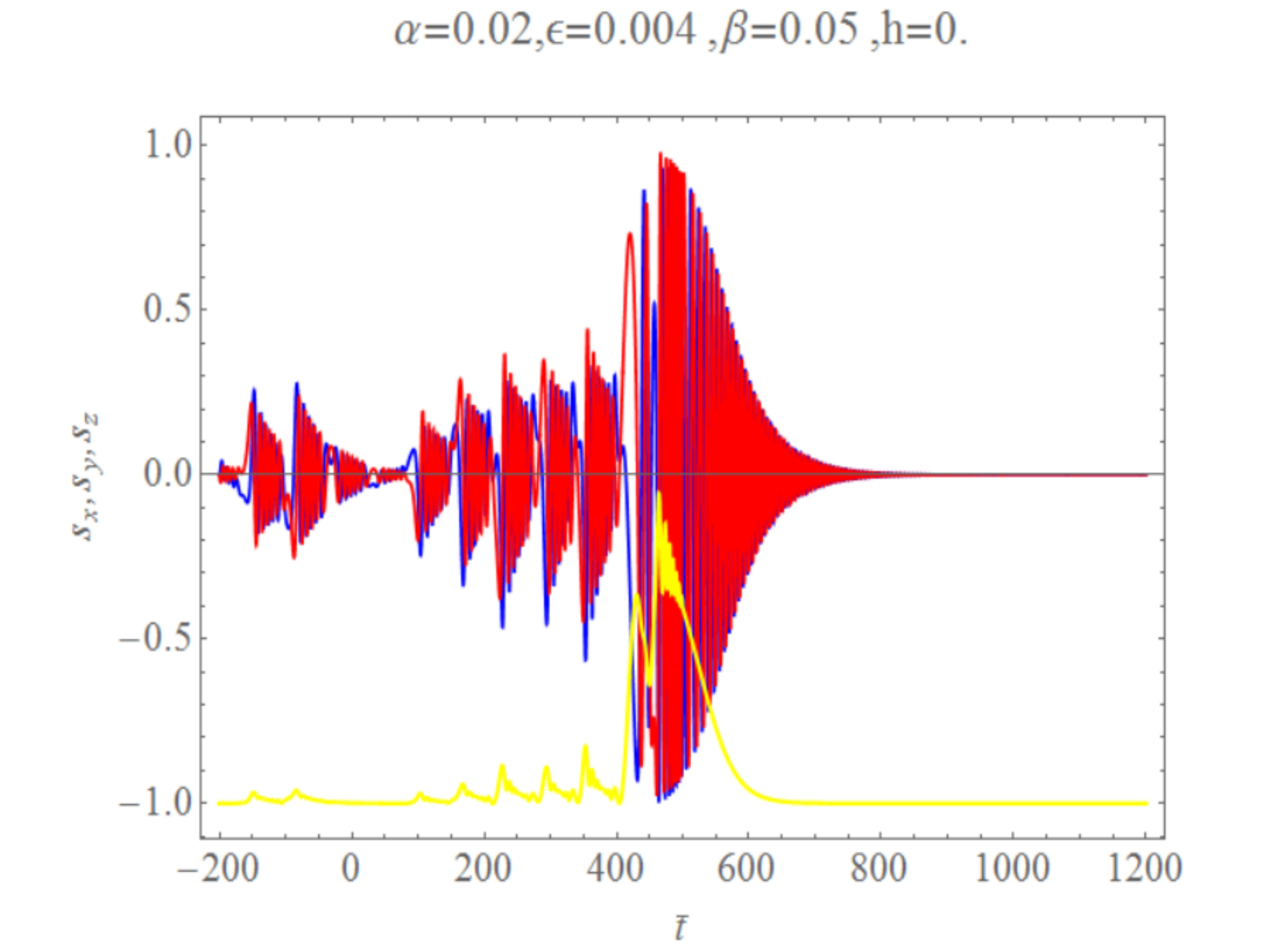}{a}
\includegraphics[width=85mm]{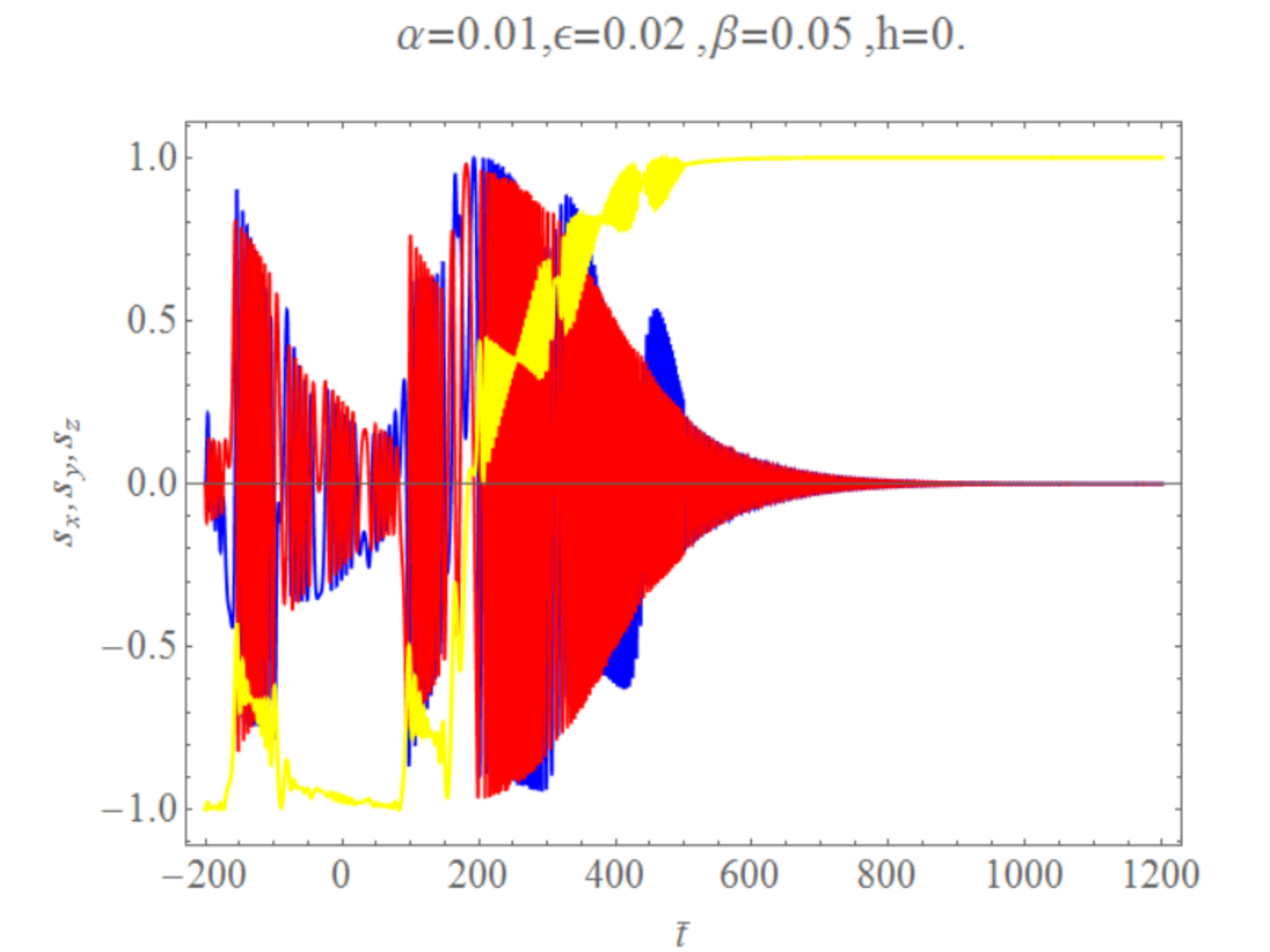}{b}
\caption{Color online: Examples of magnetization dynamics at $h = 0$ generated by a pulse of SAW with a reduced time-dependent frequency $\bar{\omega} = 1 + 0.9\cos(0.05 \bar{t})$ for two different sets of the reduced amplitude $\epsilon$ and damping parameter $\alpha$. Time dependence of $s_x,s_y,s_z$ is shown in red, blue, and yellow respectively. (a) The dynamics at $\alpha = 0.02$ and $\epsilon = 0.004$ that does not result in the magnetization switching. (b) Magnetization switching at $\alpha = 0.01$ and $\epsilon = 0.02$. The SAW power is turned off after the switching has occurred.}
\label{switching}
\end{figure}
The Landau-Lifshitz equation for the spin ${\bf S}$ of the nanomagnet in the chosen coordinate frame is
\begin{equation}
\frac{d\mathbf{{S}}}{dt}=\mathbf{S}\times{\gamma\bf H}_{eff} -\frac{\alpha}{S}\mathbf{S}\times\left(\mathbf{S}\times{\gamma\bf H}_{eff}\right),
\end{equation}
where $\alpha \ll 1$ is a dimensionlesss damping parameter. 
Dividing it by $S$ and by $\omega_{FMR}^{(0)} = 2\gamma Km$, which is the FMR frequency at $H = 0$, and introducing dimensionless
\begin{equation}
{\bf s} = \frac{\bf S}{S}, \quad \bar{t} = \omega_{FMR}^{(0)} t, \quad \bar{\omega} = \frac{\omega}{\omega_{FMR}^{(0)}}, \quad h = \frac{H}{H_A}
\end{equation}
we get
\begin{equation}
\frac{d\mathbf{{s}}}{d\bar{t}}=\mathbf{s}\times {\bf h}_{eff} -\alpha \mathbf{s}\times\left(\mathbf{s}\times {\bf h}_{eff}\right),\label{eq:LLE-Rotating_frame}
\end{equation}
where 
\begin{equation}
{\bf h}_{eff} = \epsilon(h + \bar{\omega} + \dot{\bar{\omega}} \bar{t})\hat{\bf x} + (s_z + h + \bar{\omega})\hat{\bf z}.
\end{equation}
At small damping, $\alpha \ll 1$, one obtains
\begin{eqnarray}\label{eq-s-alpha}
\frac{ds_x}{d\bar{t}} & = & s_y(s_z + h + \bar{\omega}) - \alpha s_zs_x(s_z + h)  \nonumber \\
\frac{ds_y}{d\bar{t}} & = & - s_x(s_z + h + \bar{\omega}) + s_z\epsilon (h + \bar{\omega} + \dot{\bar{\omega}} \bar{t}) \nonumber \\
& - & \alpha s_zs_y(s_z + h) \\
\frac{ds_z}{d\bar{t}} & = & -s_y\epsilon (h + \bar{\omega} + \dot{\bar{\omega}} \bar{t}) +\alpha (s_x^2 + s_y^2)(s_z + h)  \nonumber
\end{eqnarray}
for the components of spin ${\bf s}$ and the following equations for the spherical angles, ${\bf s} = (\sin\theta cos\phi, \sin\theta \sin \phi, \cos\theta)$:
\begin{eqnarray}\label{eq-angles-alpha}
&& \frac{d\theta}{d\bar{t}} = \epsilon (h + \bar{\omega} + \dot{\bar{\omega}} \bar{t})\sin\phi - \alpha \sin \theta (\cos\theta + h) \\
&& \frac{d\phi}{d\bar{t}} = - (\cos\theta + h + \bar{\omega})  + \epsilon (h + \bar{\omega} + \dot{\bar{\omega}} \bar{t})\cos\phi \cot\theta. \nonumber
\end{eqnarray}
In deriving Eqs. (\ref{eq-s-alpha}) and (\ref{eq-angles-alpha}) we dropped $\bar{\omega}$ in the terms proportional to $\alpha$ because, unlike spin precession, the damping does not depend on the observer and, thus, should not be affected by the switching to the coordinate frame rotating at the angular velocity $\bar{\omega}$. 

\begin{figure}
\includegraphics[width=100mm]{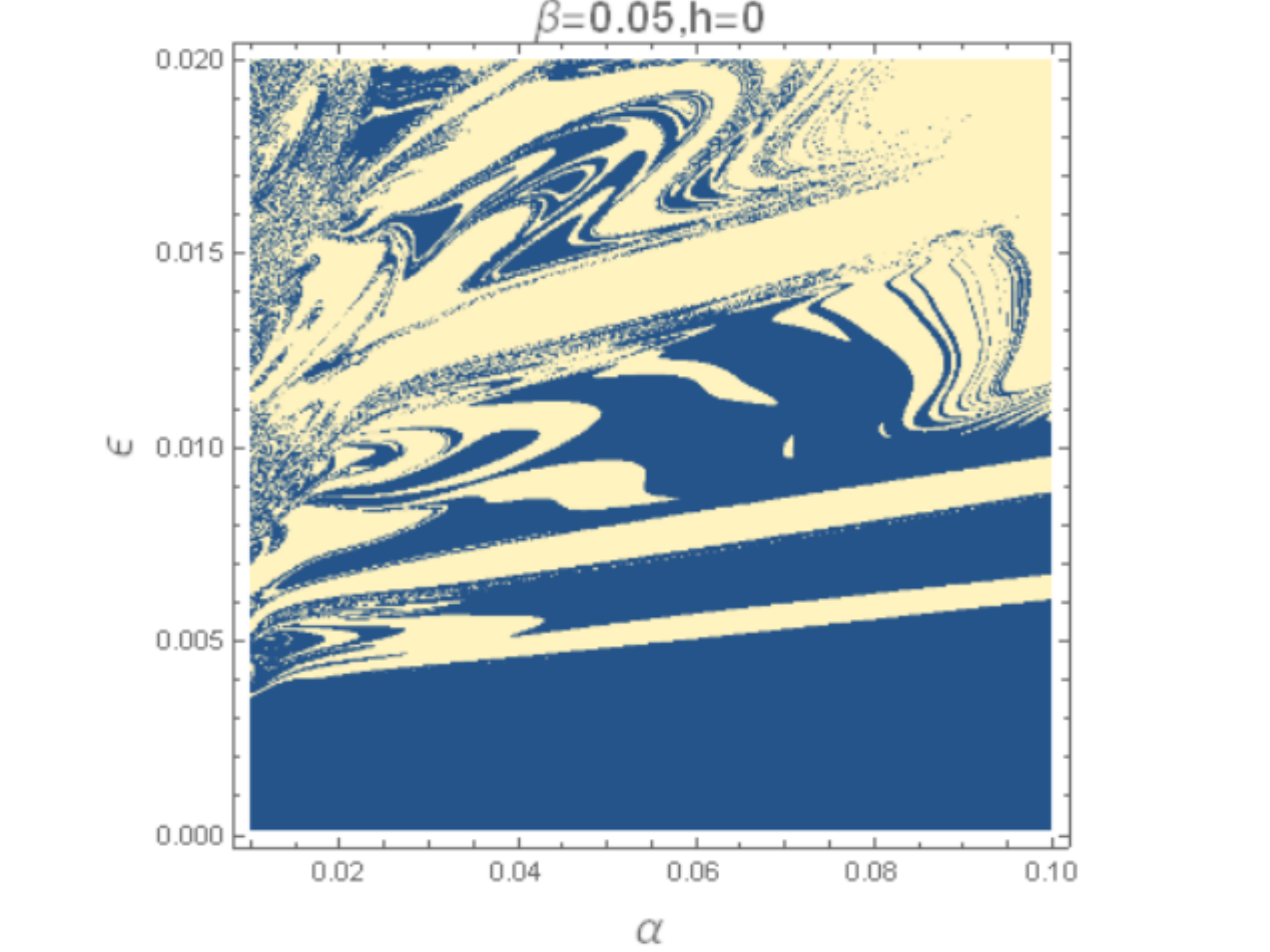}
\caption{Color online: Reduced amplitude vs damping ($\epsilon,\alpha$) phase diagram at $h = 0$ for the reversal of the magnetic moment generated by a pulse of SAW with a reduced time-dependent frequency $\bar{\omega} = 1 + 0.9\cos(0.05 \bar{t})$. The area of magnetization switching is shown by light color.}
\label{PhaseD}
\end{figure}
In quantum terms, to provide switching, the spin ${\bf S}$ of the nanomagnet should be absorbing polarized phonons of spin 1 \cite{garchu-PRB2015} in such a manner that $S_z$ changes from $S_z = -S$ to $S_z = 0$, after which ${\bf S}$ relaxes to $S_z = S$. Classically, the switching requires the magnetization to climb the anisotropy barrier towards $m_z = 0$ while precessing at the frequency $\gamma[H + 2Km_z(t)]$. The precession must be generated by SAW of time-dependent frequency $\bar{\omega} = f(\beta \bar{t})$ where $\beta$ represents the speed of the frequency change. In our model the function $f$ must satisfy the condition $f \neq 0$ to provide a finite small amplitude of SAW at the minimal frequency in accordance with the condition of constant power, $\omega u_0 = {\rm const}$, imposed on the sources of SAW. 

The above equations have been solved numerically for various values of parameters $\epsilon$, $\alpha$ and $h$. The parameter $\epsilon$ is of the order of the ratio of the amplitude of SAW and the wavelength of sound, $\epsilon \sim u_0/\lambda \ll 1$. The damping parameter for a nanoparticle embedded in a solid is typically in the range $0.01 < \alpha < 0.5$ \cite{Kalmykov-PRB2010}. The parameter $h$ in the above equations is the external magnetic field in units of the anisotropy field. Below we present numerical results for the case of $h = 0$ which has the highest practical interest: Magnetization switching by SAW unassisted by the external field. 

Figs.\ \ref{switching} and \ref{PhaseD} illustrate the possibility of the reversal of the magnetic moment of a nanoparticle by a pulse of SAW of time-dependent amplitude and the frequency of the form $\bar{\omega} = 1 + a\cos(\beta \bar{t})$.  For illustration we choose $a = 0.9$ and $\beta = 0.05$. At a given value of the Gilbert damping $\alpha$ a threshold SAW power, $P \propto \epsilon^2$, is required for the reversal, see Fig.\ \ref{PhaseD}. Some stochasticity, however, is present above the threshold. It can be traced to the strong non-linearity of Eqs.\ (\ref{eq-s-alpha}) and (\ref{eq-angles-alpha}). Nevertheless, there are large areas of deterministic switching. In practical terms this means that one will have to carefully choose the range of parameters in a device providing the magnetization reversal by SAW. 

In Conclusion, we have demonstrated that the magnetic moment of a single-domain particle embedded into the surface of a solid can be reversed by surface acoustic waves in the FMR frequency range. The physics behind this phenomenon is similar to the switching of the magnetic moment by the ac electromagnetic field. A particle subjected to SAW undergoes rotational oscillations due the shear component of the elastic deformation generated by SAW. In accordance with the Larmor theorem the effect of these rotations in the coordinate frame of the particle is the effective ac magnetic field acting on the magnetic moment. The main difference from the case of the real ac magnetic field is that the frequency and the amplitude of the effective magnetic field due to SAW are not independent. As long as the temperature of the system is small compared to the blocking temperature for superparamagnetic behavior, the effect of finite temperature on the dynamics of the magnetization should be weak. We derived the equations of motion that govern the dynamics of the magnetic moment driven by SAW of time-dependent frequency and solved them numerically in a practical range of parameters. The switching phase diagram has been constructed that shows deterministic switching in certain ranges of parameters and more stochastic switching in other ranges.  While our suggestion is for the switching of the magnetic moment in an individual nanomagnet, less demanding experiments indicating the presence of the effect can be conducted on the array of nanoparticles deposited on a surface and magnetized in a certain direction. Acceleration of the magnetic relaxation due to SAW in the FMR frequency range in the absence of the magnetic field would constitute experimental evidence of the effect. 

The authors acknowledge useful advice from Prof. Dmitry Garanin on how to increase computational speed. This research has been supported by the U.S. National Science Foundation through Grant No. DMR-1161571.

\end{document}